\documentclass[twocolumn,floatfix,aps]{revtex4}
\usepackage{epsfig}

\def\lp {\left( }
\def\rp {\right) }
\def\lb {\left[ }
\def\rb {\right] }
\def\lc {\left\{ }
\def\rc {\right\} }

\def\nn {\nonumber}

\def\beq{\begin{equation}}
\def\eeq{\end{equation}}
\def\bea{\begin{eqnarray}}
\def\eea{\end{eqnarray}}
\def\ni{\noindent}

\def\ub {\bar u}

\def\Qs {\not\!\!Q}

\def\Pb {\bar{\Pi}}

\def\mb {M}

\def\sb {\bar{s}}

\def\g{\gamma}

\def\L {\Lambda}
\def\m{\mu}
\def\n{\nu}

\def\O{\Omega}
\def\p{\pi}
\def\P{\Pi}

\def\s{\sigma}
\def\S{\Sigma}

\def\bp {\mbox{\boldmath $p$}}

\begin{document}

\title{$\pi\Xi$ phase shifts and $CP$ Violation in  
${ \Omega\rightarrow\pi\Xi}$ Decay}

\author{C.C. Barros Jr.}

\affiliation{Nuclear Theory and Elementary Particle Phenomenology Group\\
Instituto de F\'{\i}sica, Universidade de S\~{a}o Paulo,\\
C.P. 66318, 05315-970, S\~ao Paulo, SP, Brazil}

\date{\today}

\begin{abstract}
In the study of $CP$ violation signals in  $\O \rightarrow\pi\Xi$ 
nonleptonic decays, the strong $J$=3/2 $P$ and $D$ 
phase shifts for the $\pi\Xi$
final-state interactions are needed.
These phases are calculated using an effective Lagrangian model, that 
considers 
 $\Xi$, $\Xi^*$(1530), $\rho$ and the $\sigma$-term, in the 
intermediate states. The $\sigma$-term is calculated in terms of the scalar 
form factor of the baryon.
\end{abstract}

\maketitle

\vspace{5mm}

\section{introduction}

In the search for physics beyond the standard model, the observation of $CP$
violation  could be a very usefull tool. Many models shows $CP$ nonconservation
effects, as, for example, the superweak model \cite{sw}, the Kobayashi-Maskawa
 model \cite{KM}, the penguin model \cite{pen} or the Weinberg-Higgs model 
\cite{WH}.

Today, there are three systems where  $CP$ violation has been observed. 
The first one was the 
$K_L^0\rightarrow \pi^+\pi^-$ decay, where it was shown \cite{dec}
that $K_2^0$ is not a pure eigenstate of $CP$, and then, 
the parameter $\epsilon$ is nonzero. Later, direct  $CP$ violation 
in $K \rightarrow \pi \pi$ decays has been observed. 
Very recently,  $CP$ violation has also been observed in the  
$B \rightarrow J/ \Psi K_S$ and other related modes (for a review, see 
\cite{nir}).

In 1957, Okubo \cite{OK} noted that $CP$ violation could cause differences in 
the branching ratios of the $\S$ and ${\overline\S}^+$ decays. Pais \cite{PA}
extended this proposal also to $\L$ and $\overline\L$ decays.
In more recent works the $CP$ violation signs were also investigated using 
nonleptonic hyperon decays \cite{don}-\cite{meis} 
where $\Xi$ decays were stiudied too. 
In \cite{ta2} the study has been made in the framework of the standard model,
and in \cite{xh1}, \cite{xh2}, new physics was considered.
At the experimental level, there are experiments searching for $CP$ violation
in hyperon decays \cite{luk}.
  
The nonleptonic $\O$ decays has only been studied in \cite{ta3} where the 
the strong phases were estimated at leading order in
 heavy-baryon chiral perturbation theory. 
The aim of this paper is to calculate the $\Xi\pi$ strong phases
using an effective Lagrangian model, without the heavy-baryon
approximation, and their
effects to the asymmetry parameters in the $\Omega^-$ decays.

 The size of the $CP$ violation depends on the final-state 
strong interaction between the produced particles. So, in order to perform the
calculations, the strong phase shifts are needed. At the moment, the  
situation of the
$\Lambda\rightarrow\pi N$ and $\Sigma\rightarrow\pi N$  decays is very 
confortable, the $\pi N$ strong phase shift analysis is very well known 
experimentally \cite{PS}. At the theoretical level, this system is 
very well described, and, at least at low energies, the chiral perturbation 
theory is very precise \cite{LG}, \cite{PiN}.
 
However, in the decays that produce hyperons, the situation is not so good, 
 no experimental data is available to the $\pi Y$ interactions. In fact, some 
information can be obtained in the study of hyperonic atoms (see, for example,
\cite{LOI}, where the $\pi\L\S$ coupling constant is estimated), but it is not
enough to fully understand the $\pi \L$ interaction. Thus, to investigate the 
$\pi Y$ interactions, the only way is to use a model.

As it was said, the chiral perturbation theory is very accurate when applied 
to the $\pi N$ interactions, so, we hope that it works in the $\pi Y$ system 
too.   
In \cite{Kam}-\cite{meis}, \cite{keis}
that was done to the $\Xi\rightarrow\pi\L$ decay, 
and  the calculated phases were very small. 

In this work, the $\O\rightarrow\pi\Xi$ is studied, and the strong phase shifts
for $\pi \Xi$ interactions are calculated 
using the model presented in \cite{BH}.
In \cite{BH}, chiral lagrangians are used, describing processes with  
 $\Xi$, $\Xi^*$ 
and  $\rho$  in the intermediate states. The 
$\sigma$-term is also included, but not only as a parametrization (as it was 
done in \cite{BH}), but relating it with the scalar form factor $\sigma(t)$,
based in the results of \cite{MAN}, \cite{CM}. 

This paper will show the following contents: In section II it will be shown the
$\Omega^-$ decay and how to calculate the observables. In section III we will 
calculate the  phase shifts in the $\pi\Xi$ interactions. The results and 
conclusions are in section IV.

\section{Nonleptonic ${\bf \Omega^-}$ Decay}

In the  $\Omega^-$ ($J^p={3\over 2}^+$) decays, the transitions are of the form
\beq
{\rm spin}\ 3/2 \rightarrow {\rm spin}\ 0 + {\rm spin}\ 1/2 \  \  , \nonumber
\eeq

\ni
and the contributing phases are the
 $J={3\over 2}$ $P$(parity conserving) and $D$(parity violating)
 waves in the $\Omega$ rest frame. The $\Delta S=1$ $\pi\Xi $ nonleptonic
decays are
$\Omega^-\rightarrow\Xi^0\pi^-$ and $\Omega^-\rightarrow\Xi^-\pi^0$.

The experimental observables are the total rate $\Gamma$, and the asymmetry 
parameters, that can be written as
\beq
\alpha=2\ {\rm Re}(P^*D)/(|P|^2+|D|^2)
\eeq
\beq
\beta=2\ {\rm Im}(P^*D)/(|P|^2+|D|^2)
\eeq
\beq
\gamma=(|P|^2-|D|^2)/(|P|^2+|D|^2)
\eeq

\noindent
and obeys the relation
\beq
\alpha^2+\beta^2+\gamma^2=1  \  \  .
\eeq

\noindent
For antihyperons decays the expressions are
\beq
{\overline\alpha}=2\ {\rm Re}({\overline P}^*\ {\overline D})/
(|{\overline P}|^2+|{\overline D}|^2)
\eeq
\beq
{\overline\beta}=2\ {\rm Im}({\overline P}^*\ {\overline D})/
(|{\overline P}|^2+|{\overline D}|^2) \  \  .
\eeq

\noindent
The $P$ and $D$ amplitudes can be parametrized as
\beq
P=\sum_Ia_{2I}|P_{2I}|e^{i(\delta_P^{2I}+\phi_P^{2I})}
\label{2.12}
\eeq
\beq
D=\sum_Ia_{2I}|D_{2I}|e^{i(\delta_D^{2I}+\phi_D^{2I})}  \  \  ,
\label{2.13}
\eeq

\noindent
where $I$ is the isospin state, $\delta_l^{2I}$ are the strong phase shifts and
$\phi_l^{2I}$ are the weak $CP$ violating phases. The respective $CP$ 
conjugated amplitudes are
\beq
{\overline P}=\sum_Ia_{2I}|P_{2I}|e^{i(\delta_P^{2I}-\phi_P^{2I})}
\eeq
\beq
{\overline D}=-\sum_Ia_{2I}|D_{2I}|e^{i(\delta_D^{2I}-\phi_D^{2I})}  \  \  ,
\eeq

\noindent
Using  eq. (\ref{2.12}), (\ref{2.13}) 
 for the $\Omega^-\rightarrow\Xi^0\pi^-$ decay we have
\beq
P(\Omega_-^-)=-\sqrt{2\over 3}P_{1}e^{i(\delta_P^1+\phi_P^1)}+
\sqrt{1\over 3}P_{3}e^{i(\delta_P^3+\phi_P^3)}
\eeq
\beq
D(\Omega_-^-)=-\sqrt{2\over 3}D_{1}e^{i(\delta_D^1+\phi_D^1)}+
\sqrt{1\over 3}D_{3}e^{i(\delta_D^3+\phi_D^3)}
\eeq

\noindent
and for $\Omega^-\rightarrow\Xi^-\pi^0$
\beq
P(\Omega_0^-)=\sqrt{1\over 3}P_{1}e^{i(\delta_P^1+\phi_P^1)}+
\sqrt{2\over 3}P_{3}e^{i(\delta_P^3+\phi_P^3)}
\eeq
\beq
D(\Omega_0^-)=\sqrt{1\over 3}D_{1}e^{i(\delta_D^1+\phi_D^1)}+
\sqrt{2\over 3}D_{3}e^{i(\delta_D^3+\phi_D^3)}  \  \  .
\eeq

In the limit of $CP$ conservation, the $CP$ asymmetry parameters
\beq
A={\alpha+{\overline \alpha}\over \alpha-{\overline \alpha}}
\eeq

\ni
and
\beq
B={\beta+{\overline \beta}\over \beta-{\overline \beta}}
\eeq

\ni 
vanish, since $\alpha$=-${\overline \alpha}$ and $\beta$=-${\overline \beta}$. 
In hyperon decays,
the $\Delta I$=3/2 amplitudes are much smaller then the
$\Delta I$=1/2, then, in the first order in  $\Delta I$=3/2 amplitudes, 
\bea
A(\Omega_-^-)&=&-{\rm tan}(\delta_P^1-\delta_D^1){\rm tan}(\phi_P^1-\phi_D^1)
\biggl\{  1 + 
\nonumber \\
&&
+{1\over \sqrt{2}}{P_3\over P_1}\lb 
{\cos(\delta_P^3-\delta_D^1)\cos(\phi_P^3-\phi_D^1)\over
\cos(\delta_P^1-\delta_D^1)\cos(\phi_P^1-\phi_D^1)} \right. \nn \\
&& \left.
-
{\sin(\delta_P^3-\delta_D^1)\sin(\phi_P^3-\phi_D^1)\over
\sin(\delta_P^1-\delta_D^1)\sin(\phi_P^1-\phi_D^1)}
   \rb  \nonumber \\
&&
+{1\over \sqrt{2}}{D_3\over D_1}\lb 
{\cos(\delta_P^3-\delta_D^1)\cos(\phi_P^3-\phi_D^1)\over
\cos(\delta_P^1-\delta_D^1)\cos(\phi_P^1-\phi_D^1)} \right. \nn \\
&& \left.
-{\sin(\delta_P^3-\delta_D^1)\sin(\phi_P^3-\phi_D^1)\over
\sin(\delta_P^1-\delta_D^1)\sin(\phi_P^1-\phi_D^1)}
   \rb \biggr\}
\label{2.10} \\
B(\Omega_-^-)&=&{\rm cot}(\delta_P^1-\delta_D^1){\rm tan}(\phi_P^1-\phi_D^1)
\biggl\{  1 + 
\nonumber \\
&&
+{1\over \sqrt{2}}{P_3\over P_1}\lb 
{\sin(\delta_P^3-\delta_D^1)\cos(\phi_P^3-\phi_D^1)\over
\sin(\delta_P^1-\delta_D^1)\cos(\phi_P^1-\phi_D^1)} \right. \nn \\
&& \left.
-{\cos(\delta_P^3-\delta_D^1)\sin(\phi_P^3-\phi_D^1)\over
\cos(\delta_P^1-\delta_D^1)\sin(\phi_P^1-\phi_D^1)}
   \rb  \nonumber \\
&&
+{1\over \sqrt{2}}{D_3\over D_1}\lb 
{\sin(\delta_P^3-\delta_D^1)\cos(\phi_P^3-\phi_D^1)\over
\sin(\delta_P^1-\delta_D^1)\cos(\phi_P^1-\phi_D^1)} \right. \nn \\
&& \left.
-{\cos(\delta_P^3-\delta_D^1)\sin(\phi_P^3-\phi_D^1)\over
\cos(\delta_P^1-\delta_D^1)\sin(\phi_P^1-\phi_D^1)}
   \rb \biggr\}
\label{2.11}
\eea

\ni
and similar expressions for the $\Omega\rightarrow \pi^0\Xi^-$ decays 
(replacing the factors $1/\sqrt{2}$ for $-\sqrt{2}$ in the expressions 
(\ref{2.10}), (\ref{2.11})).
At leading order, 
\beq
A(\Omega_-^-)=A(\Omega_-^0)=
-{\rm tan}(\delta_P^1-\delta_D^1){\rm tan}(\phi_P^1-\phi_D^1)  \  \  ,
\label{1.30}
\eeq

\ni
and
\beq
B(\Omega_-^-)=B(\Omega_-^0)=
{\rm cot}(\delta_P^1-\delta_D^1){\rm tan}(\phi_P^1-\phi_D^1)  \  \  .
\label{1.31}
\eeq

\noindent

In the next section we will calculate the phase shifts $\delta_l^{2I}$, that
are needed to estimate $A$ and $B$.

\section{Low energy ${\pi\Xi}$ interaction}

In order to describe the low energy $\pi\Xi$ interaction, a reliable 
way is to use effective Lagrangians, as it was done in a previous 
work \cite{BH}. A very important feature of this model is to allow the 
inclusion of spin 3/2 ressonances in the intermediate states. 
In the low energy $\pi^+ P$ interactons, for
example, the $\Delta$(1232) dominates almost completely the total cross 
section (when $\sqrt{s}$ is near the $\Delta$ mass). 
Consequently, it is expected that in  some reactions of $\pi Y$ 
scattering, the same behaviour will occur \cite{BH}. The lagrangians to be 
considered are
\begin{eqnarray}
{\cal{L}}_{\Xi\pi\Xi} &=& {g_{\Xi\pi\Xi}\over 2m_\Xi}\lbrack {\overline\Xi}
\gamma_\mu \gamma _{5} \vec \tau\Xi \rbrack .\partial^\mu \vec \phi  \\
{\cal{L}}_{\Xi\pi\Xi^*} &=& g_{\Xi\pi\Xi^*} \lbrace {{\overline{\Xi}^*}^\mu}
\lbrack g_{\mu\nu} - (Z+{1\over 2})\gamma_\mu \gamma _\nu  
\rbrack \vec\tau\Xi \rbrace .\partial ^\nu \vec \phi + H.c. \nonumber \\
&&  \label{3.1} \\
{\cal{L}}_{\Xi\rho\Xi} &=& {\gamma_0\over 2}\lbrack {\overline \Xi}\gamma_\mu 
\vec\tau\Xi \rbrack \vec\rho^\mu \nonumber \\
&&
+ {\gamma_0\over 2}\lbrack {\overline \Xi} ({\mu_{\Xi^0}-\mu_{\Xi^-}
\over 4m_\Xi})i
\sigma_{\mu\nu}\vec\tau\Xi \rbrack .(\partial ^\mu \vec{\rho^\nu} - 
\partial ^\nu \vec{\rho^\mu}) \\
{\cal{L}}_{\rho\pi\pi} &=& \gamma_0\vec\rho_\mu.(\vec\phi\times
\partial^\mu \vec \phi) \nonumber \\
&& - {\gamma_0\over 4m_\rho^2}(\partial _\mu 
\vec\rho_\nu - \partial_\nu\vec\rho_\mu).(\partial^\mu \vec\phi\times
\partial^\nu \vec \phi)
   \   \      , 
\end{eqnarray}

\noindent
where $\Xi$, $\Xi^*$, $\vec\phi$ and $\vec\rho$ are the cascade, the resonance 
$\Xi^*$(1530), the pion and the rho fields. $Z$ is the off-shell parameter 
\cite{PiN},
$\mu_{\Xi^0}$ and $\mu_{\Xi^-}$ are the magnetic moments.

The lagrangians are almost the same as the $\pi N$ ones \cite{PiN}, because 
the $\pi N$ 
system is very similar to the $\pi \Xi$. $N$ and $\Xi$ are particles 
with isospin 1/2, the only difference is that $\Delta$(1232) has isospin 3/2 
and $\Xi^*$(1530), 1/2, so, a $\vec\tau$ matrix is included in (\ref{3.1}).  

The spin 3/2 propagator for a mass $M$ particle, is then 
\bea
G^{\m\n}(p)  &=& - \; \frac{(\not\!\!{p}+M)}{p^2-M^2}\lp g^{\m\n}
 -\frac{\g^\m\g^\n}{3}
 \right. \nn \\
& &\left.
-\frac{\g^\m p^\n}{3M} + \frac{p^\m \g^\n}{3M} - \frac{2 p^\m p^\n}{3M^2}\rp \;.
\label{2.3}
\eea

\begin{figure}[hbtp]
\centerline{
\epsfxsize=80.mm
\epsffile{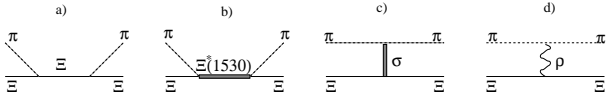}}
\caption{Diagrams to $\pi\Xi$ Interaction.}
\end{figure}


The contributing diagrams are shown in the Fig. 1 (we show only the direct
diagrams, but in the calculations, the crossed diagrams are also included). 
The scattering matrix will have the general form
\bea
T_{\pi\Xi}^{ba}&=& \overline{u}(\vec{p}\prime )\lbrace \lbrack A^+ + 
{(\not\!k + \not\!k')\over 2}B^+\rbrack \delta_{ba} \nn \\
&&
+ \lbrack  A^- + 
{(\not\!k + \not\!k')\over 2}A^-
\rbrack i\epsilon_{bac}\tau^c\rbrace u(\vec p)      \    \     ,
\eea

\ni
where $k$ and $k'$ are the initial and final $\pi$ momenta. 
Calculating the amplitudes from the diagrams,  the contributions from Fig. 
1(a) (intermediate $\Xi$) are
\begin{eqnarray}
& &A_\Xi^+ = {g_{\Xi\pi\Xi}^2\over m_\Xi} \nn  \\
& &A_\Xi^-=0 \nn  \\  
& &B_\Xi^+ = g_{\Xi\pi\Xi}^2\lb {1\over u-m_\Xi^2}-  
{1\over s-m_\Xi^2}\rb
 \nn \\ 
& &B_\Xi^- =-{g_{\Xi\pi\Xi}^2\over 2 m_\Xi}
-g_{\Xi\pi\Xi}^2\lb {1\over u-m_\Xi^2}+  
{1\over s-m_\Xi^2}\rb  \  \  .    
\end{eqnarray}

\noindent
Fig 1(d), the $\rho$ exchange, gives
\begin{eqnarray}
& &A_\rho^+ = B_\rho^+ = 0 \nn  \\ 
& &A_\rho^- = -{\gamma_0^2\over m_\rho^2}(\mu_{\Xi^0} - \mu_{\Xi^-})\nu
{1 - {t/ 4m_\rho^2}\over 1 - {t/ m_\rho^2}} \nn  \\
& &B_\rho^- = {\gamma_0^2\over m_\rho^2}(1 + \mu_{\Xi^0} - \mu_{\Xi^-})
{1 - {t/ 4m_\rho^2}\over 1 - {t/ m_\rho^2}}  \  \  . 
\end{eqnarray}

\noindent
The contribution from Fig. 1(b), the interaction with the intermediate 
$\Xi^*$, is
\begin{eqnarray}
A_{\Xi^*}^+ &=& {g_{\Xi\pi\Xi^*}^2\over 3m_\Xi}\lbrace {\nu_r \over 
\nu_r^2 - \nu^2}\hat A \nonumber \\
&&
- {m_\Xi^2+m_\Xi m_{\Xi^*}\over m_{\Xi^*}^2}(2m_{\Xi^*}^2
m_\Xi m_{\Xi^*}-m_\Xi^2+2\mu^2) \nonumber \\
& & +{4m_\Xi\over m_{\Xi^*}^2}\lbrack (m_\Xi+m_{\Xi^*})Z + 
(2m_{\Xi^*}+m_\Xi)Z^2 \rbrack k.k' \rbrace \nonumber  \\
&&   \\
A_{\Xi^*}^- &=& {g_{\Xi\pi\Xi^*}^2\over 3m_\Xi}\lbrace {\nu \over 
\nu_r^2 - \nu^2}\hat A + 
{8m_\Xi^2\nu\over m_{\Xi^*}^2}\lbrack (m_\Xi+m_{\Xi^*})Z \nonumber \\
&&
+ (2m_{\Xi^*}+m_\Xi)Z^2 \rbrack \rbrace   \\
B_{\Xi^*}^+ &=& {g_{\Xi\pi\Xi^*}^2\over 3m_\Xi}\lbrace {\nu \over 
\nu_r^2 - \nu^2}\hat B - {8m_\Xi^2\nu Z^2\over m_{\Xi^*}^2} \rbrace   \\
B_{\Xi^*}^- &=& {g_{\Xi\pi\Xi^*}^2\over 3m_\Xi}\lbrace {\nu_r \over 
\nu_r^2 - \nu^2}\hat B 
-{4m_\Xi\over m_{\Xi^*}^2}\lbrack (2m_\Xi^2 \nonumber \\
&&
+2m_\Xi m_{\Xi^*}-2\mu^2)Z
 + (2m_\Xi^2+4m_\Xi m_{\Xi^*})Z^2 \rbrack \nonumber \\
&& 
+  {(m_\Xi+m_{\Xi^*})^2 \over m_{\Xi^*}^2}
 - {4m_\Xi Z^2\over m_{\Xi^*}^2}k.k' \rbrace \  \   ,
\end{eqnarray}

\noindent
with
\begin{eqnarray}
\hat A &=& {(m_{\Xi^*}+m_\Xi)^2-\mu^2\over 2m_{\Xi^*}^2}\lbrack 
2m_{\Xi^*}^3-2m_\Xi^3-
2m_\Xi m_{\Xi^*}^2  \nonumber \\
&&
-2m_\Xi^2m_{\Xi^*}+\mu^2(2m_\Xi-m_{\Xi^*})  \rbrack  + 
\nonumber \\
& &+ {3\over 2}(m_\Xi+m_{\Xi^*})t \\
\hat B &=& {1\over 2m_{\Xi^*}^2}\lbrack (m_{\Xi^*}^2-m_\Xi^2)^2
-2m_\Xi m_{\Xi^*}(m_\Xi^*+m_\Xi)^2 \nonumber \\
&&
+6\m^2m_\Xi (m_\Xi^*+m_\Xi)
-2\mu^2(m_\Xi^*+m_\Xi)^2+\mu^4\rbrack + {3\over 2}t \ ,  \nonumber \\
&& \label{3.2} 
\end{eqnarray}

\ni
 where $\mu$ is the pion mass, and $\n$ and $\n_r$ are defined in the 
appendix A.
One must remark that eq. (\ref{3.2}) is different from the expression of 
\cite{BH}, 
where there was a mistake. The correct expression is presented here.

In \cite{BH}, \cite{PiN} the $\sigma$ term (diagram 1.c) was simply considered 
as a   parametrization 
\begin{eqnarray}
A_\sigma &=& a+bt   \nonumber  \\
B_\sigma &=& 0 \   \   .
\end{eqnarray}

\ni
In fact, the $\sigma$ term represents the exchange of a scalar isoscalar 
system in the $t$-channel.
 This contribution is related to the scalar form factor of the
baryon, and at large distances is dominated by triangle diagrams (Figure 2)
involving the exchange of 2 pions  \cite{LW}. In the $\pi \Xi$ interaction, 
this 
contribution is associated with two triangle diagrams, with $\Xi$ and $\Xi^*$ 
intermediate states, as it was calculated in \cite{CM}.

\begin{figure}[hbtp]
\centerline{
\epsfxsize=90.mm
\epsffile{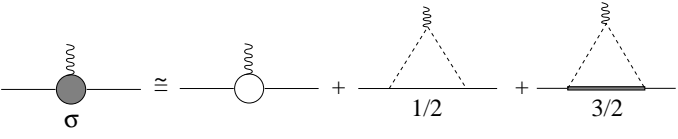}}
\caption{ The scalar form factor (grey blob)  receives contributions from tree
 interactions (white blob)
and triangle diagrams with spin $1/2$ and $3/2$ intermediate states.}
\end{figure}

The scalar form factor for a spin $1/2$ baryon B is defined as 
$ <B(p') | - {\cal L}_{sb} | B(p) > \equiv \s(t) \; \ub(\bp') \;u(\bp) $,
where ${\cal L}_{sb}$ is the chiral symmetry breaking lagrangian.

The contribution of an intermediate particle of spin $s$ and mass $M$ to the scalar form factor is given by
\bea
&& \s_s(t;M)\; \ub\;u = - \m^2 \lp \frac{g}{2 m}\rp^2 \lp T_a^\dagger T_a \rp
\nn \\
&&
\times \int  \frac{d^4Q}{(2\pi)^4}\; 
\frac{[\ub\; \Lambda_s \;u]}{[ (Q\!-\!q/2) ^2\! -\!\m^2] 
[(Q\!+\!q/2)^2 \!-\!\m^2]} \;,
\label{2.4}
\eea

\ni
where $Q$ is the average of the internal pion momenta, $T$ is a vertex isospin
matrix  and
\bea
&& [\ub\; \Lambda_{1/2}\;u] = \ub \; \lc -(m\!+\!M) + \frac{(m\!-\!M)(m\!+\!M)^2}{s\!-\!M^2} \right. \nn \\
&& \left.
+ \lb 1 + \frac{(m\!+\!M)^2}{s\!-\!M^2} \rb \Qs  \rc \;u \;,
\label{2.6}\\[4mm]
&& [\ub\;\Lambda_{3/2} \; u] =
 - \ub \; \lc \lb \frac{1}{s\!-\!M^2}\lp (m\!+\!M) (\m^2-t/2)
\right.\right.\right. \nn\\
&& \left.\left.\left.
-\;\frac{(2 M + m) }{6 M^2}\m^4 \rp \right.\right.\nn\\
&&\left.\left.
+ \lp \frac{m^2\!-\!M^2}{s\!-\!M^2}-1 \rp \;\frac{(m\!+\!M)}{6 M^2} 
\lp (m\!+\!M) (2 M - m ) + 2\m^2 \rp\right.\right.\nn\\
&&\left.\left. 
-\;\frac{m\; (s\!-\! m^2)}{6 M^2}\rb
+\lb \frac{1}{s\!-\! M^2} \lp (\m^2-t/2)
+\frac{2 m}{3}(m\!+\!M)\right.\right.\right.\nn\\
&&\left.\left.\left.
-\frac{(m\!+\!M)\m^2}{3M} 
- \frac{\m^4}{6M^2} \rp 
+ \lp \frac{m^2\!-\!M^2}{s\!-\!M^2}-1 \rp \;\right.\right. \nn\\
&& \left.\left.
\times\frac{1}{6M^2} \lp M^2 + 2m M - m^2 + 2\m^2 \rp  
-\;\frac{s\!-\!m^2}{6M^2} \rb \Qs \rc \;u
\label{2.7}\;.
\eea

Calculating $\sigma(t)$ in the $\pi\Xi$ interaction, and
using the loop integrals $\P$ defined in appendix B, we obtain 
\bea
&& \s_{1/2}(t) = \frac{\m^2}{(4\p)^2}\lp \frac{g_{\Xi\p\Xi}}{2m_\Xi}\rp^2
  (m_\Xi\!+\!m_{\Xi^*})
\lb \P_{cc}^{(000)}- \right. \nn\\
&&\left.
\frac{m_\Xi^2\!-\!m_{\Xi^*}^2}{2m_\Xi\m}\;\P_{\sb c}^{(000)} -\;
\frac{m_\Xi+m_{\Xi^*}}{2m_\Xi}\;\P_{\sb c}^{(001)} \rb\;,
\label{2.8}\\[4mm]
&& \s_{3/2}(t) = \frac{\m^2}{(4\p)^2} \lp 
\frac{g_{\Xi\p\Xi^*}}{2m_\Xi}\rp^2 \frac{1}{6m_{\Xi^*}^2}
\lc - \lb (m_\Xi\!+\!m_{\Xi^*})^2 \right.\right. \nn\\
&&\left.\left.
\times(2m_{\Xi^*}\!-\!m_\Xi) + 2 \m^2 (m_\Xi\!+\!m_{\Xi^*})
+ (\m^2\!-\!t/2) \; m_\Xi  \rb \P_{cc}^{(000)}\right.
\nn\\[2mm]
&& \left.
- 2 \m^2 m_\Xi  \; \Pb_{cc}^{(000)}
+ \lb (m_\Xi^2\!-\!m_{\Xi^*}^2) (m_\Xi\!+\!m_{\Xi^*})^2 
(2m_{\Xi^*}\!-\!m_\Xi)\right.\right.
\nn\\[2mm]
&& \left.\left. 
+ 2\m^2 (m_\Xi\!+\!m_{\Xi^*})(m_\Xi^2\!-\!m_{\Xi^*}^2) \right.\right.
\nn\\[2mm]
&& \left.\left.
+ 6  (\m^2\!-\! t/2) m_{\Xi^*}^2 (m_\Xi\!+\!m_{\Xi^*})
 -\m^4 (2m_{\Xi^*}\!+\!m_\Xi) \rb \frac{\P_{\sb c}^{(000)}}{2m_\Xi\m}
 \right.
\nn\\[2mm]
&& \left. 
+ \lb (m_\Xi\!+\!m_{\Xi^*})^2 (4m_\Xi m_{\Xi^*}\!-\!m_\Xi^2\!-\!m_{\Xi^*}^2) 
+ 6 m_{\Xi^*}^2 (\m^2\!-\! t/2) 
\right.\right.
\nn\\[2mm]
&& \left.\left.
- 2 \m^2 (m_\Xi\!+\!m_{\Xi^*}) (2m_{\Xi^*}\!-\!m_\Xi) 
- \m^4 \rb
\frac{\P_{\sb c}^{(001)}}{2m_\Xi} \rc\;.
\label{2.9}
\eea

\ni
More details, as, for example regularization and the determination of 
$\sigma(t=0)$ can be found in \cite{CM}.

The partial wave amplitudes are obtained summing the contributions from the 
diagrams of fig. 1 and making a straightforward application of the expressions 
found in  appendix A. This can be done, for example, in the $\pi\Xi$
center-of-mass frame, 
where $\kappa$ is the momentum and $x=cos\ \theta$, where $\theta$ is
the scattering angle.
One notes that the $a_{l\pm}$ amplitudes are real, and, so, the
corresponding $S$ matrix is not unitary. To unitarize the amplitudes, 
we reinterpret them  as elements of the $K$ matrix \cite{EB}, and then
\beq
a_{l\pm}^U={a_{l\pm}\over 1-i\kappa\ a_{l\pm}}  \  \  ,
\eeq 

\ni
where $U$ means unitarized. Now the phase shifts are
\beq
\delta_{l\pm} = {\rm tg}^{-1}(\kappa\ f_{l\pm})  \  \  .
\eeq

\noindent
The parameters used are the same that were
used in \cite{BH} and are
 $m_\Xi$=1.318 GeV, $m_{\Xi^*}$=1.533 GeV, 
$\mu_{\Xi^0}=-1.25$, $\mu_{\Xi^-}=0.349$,  $g_{\Xi\pi\Xi}$=4 and
$\gamma_0^2/m_\rho^2$=$1/(2f_\p^2)$, with $f_\p$=93 MeV.
The coupling constant $g_{\Xi\pi\Xi^*}$ 
 can be calculated comparing the ressonant $\delta_P^{1}$ phase
with the Breit-Wigner expression 
\beq
\delta_{l\pm}=\tan \lb {\Gamma_0 \lp {\kappa\over\kappa_0}\rp^{2l} \over 
2( m_r- 
\sqrt{s})} \rb
\eeq
where $\kappa_0$ is the center-of-mass momentum at the peak of the resonance.
 The obtained value is 4.54 ${\rm GeV}^{-1}$.

The numerical results of the  phase shifts at $\sqrt{s}=m_{\Omega}$ are 
\beq  
\delta_{P}^{1}= -10.173^o  \  \   {\rm and}  \  \  
\delta_{D}^{1}= 0.208^o
\label{3.20}
\eeq
\beq
\delta_{P}^{3}= 0.106^o  \  \   {\rm and}  \  \  
\delta_{D}^{3}= -0.078^o  \  \  .
\label{3.21}
\eeq

\section{Summary and conclusions}

We have calculated the strong $P$ and $D$ phase shifts for the $\Omega^-$ decay
at its mass including the contributions from the diagrams of Fig. 1.
The numerical values of the phases are shown in the expressions 
(\ref{3.20}), (\ref{3.21}). The respective values calculated in \cite{ta3} are
$\delta_{P}^{1}= -12.8^o$ and $\delta_{P}^{1}= 1.1^o$, that are greater in
magnitude than the ones obtained in this work. One must remark that the 
calculations presented here have no heavy-baryon 
approximation and  the $\rho$-exchange and the
$\sigma$-term are included, 
that are the sources of the differences. 
As we can see in eq. (\ref{3.20}), (\ref{3.21}), the $D$ phases are much 
smaller than the $P$ ones, as it is expected at low energies, and can even be 
neglected in a first approximation. We expect that the same pattern occurs
 for the weak phases. 

In the other hyperon decays, the weak phases are of the order
 $\phi_P\sim 10^{-3}$ in the Weinberg-Higgs 
model and $\phi_P\sim 10^{-4}$ in the Kobayashi-Maskawa model \cite{don}. In 
\cite{ta3}, using the KM model, the estimated value was
$\phi_P\sim 10^{-3}$, which is significantly 
larger when compared with the other hyperons. 

The asymmetry parameter $A$, eq. (\ref{1.30}), in the decays 
$\Omega\rightarrow \Xi\p$ depends on the factor 
$\tan(\delta_{P}^{1}-\delta_{D}^{1})\sim$-0.18. In the 
$\Lambda\rightarrow p\p^-$ decay, the factor is 
$\tan(\delta_{S}^1-\delta_{P}^1)\sim$-0.12, and using the results of 
\cite{Kam}, for the $\p\Lambda$ phases, in the $\Xi\rightarrow\p\Lambda$
decay, $\tan(\delta_{S}-\delta_{P})\sim$0.05. One notes that the greatest 
value happens in the $\Omega$ decay (and the weak phases are also larger),
so, we conclude that the $A$ parameter must be the largest in this decay. 
On the other hand, considering 
the $B$ parameter eq. (\ref{1.31}), the $\Omega$ decay  shows a smaller
term, from   $\cot(\delta_{P}^{1}-\delta_{D}^{1})$, but the term that depends
on the weak phases is larger. So, the  $B$ parameter is probably of the same 
order of the one that appears in the other hyperon decays.
The $B$ parameter seems to be the one where the $CP$ violation would be most
evident.

\begin{acknowledgments}
I wish to tank professor Y.Hama and professor M. R. Robilotta, for many helpful
discussions.
This work was supported by FAPESP and CNPq.
\end{acknowledgments}

\appendix\section{Basic formalism}

In this paper $p$ and $p^{\prime}$ are the initial and final hyperon
4-momenta, $k$ and $k^{\prime}$ are the initial and final pion 4-momenta, so
the Mandelstam variables are 
\begin{eqnarray}
s &=& (p+k)^2=(p^{\prime}+k^{\prime})^2 \\
t &=& (p-p^{\prime})^2=(k-k^{\prime})^2 \\
u &=& (p^{\prime}-k)^2=(p-k^{\prime})^2 \ \ .
\end{eqnarray}
With these variables, we can define 
\begin{eqnarray}
\nu &=& {\frac{s-u}{4m}} \\
\nu_0 &=& {\frac{2\m^2-t}{4m}} \\
\nu_r &=& {\frac{m_r^2-m^2-k.k^{\prime}}{2m}} \ \ ,
\end{eqnarray}
where $m$, $m_r$ and $\m$ are, respectively, the hyperon mass, the 
resonance mass and the pion mass. The scattering amplitude for an isospin 
$I$ state is 
\beq
T_I=\overline{u}(\vec p\prime)\lbrace \lbrack A^I + 
{\frac{(\not\!k + \not\!k')}{2}}B^I\rbrack\rbrace u(\vec p)\ ,  
\eeq

\ni where $A_I$ and $B_I$ are calculated using the Feynman diagrams. So the 
scattering matrix is 
\beq
M_I^{ba} = {\frac{T_I^{ba}}{8\pi\sqrt{s}}} = f_I(\theta) + \vec\sigma.\hat n
g_I(\theta) = f_1^I + {\frac{(\vec\sigma .\vec k' )(\vec\sigma .\vec k)}{kk'}}f_2^I \ \ , 
\eeq
with 
\begin{eqnarray}
& &f_1^I(\theta) = {\frac{(E+m)}{8\pi\sqrt{s}}} \lbrack A_I + (\sqrt{s}%
-m)B_I\rbrack\ \ , \\
& & f_2^I(\theta) = {\frac{(E-m)}{8\pi\sqrt{s}}} \lbrack -A_I + (\sqrt{s}%
+m)B_I\rbrack \ \ ,
\end{eqnarray}
where $E$ is the hyperon energy, and
\beq
A^{1\over 2}=A^++2A^-  \  \  ,  \  \  A^{3\over 2}=A^+-A^-  \  \  , 
\eeq

\noindent
and similar expressions holds to $B^I$.
The partial-wave decomposition is done
with 
\beq
a_{l\pm} = {\frac{1}{2}}\int_{-1}^{1}\lbrack P_l(x)f_1(x) + P_{l\pm
1}(x)f_2(x)  \rbrack dx \ \ . 
\eeq

In our calculation (tree level) $a_{l\pm}$ is real. With the unitarization, 
as explained in Section III, we obtain 
\beq
a_{l\pm}^U = {\frac{1}{2i}}\lbrack e^{2i\delta_{l\pm}} -1\rbrack =
e^{i\delta_{l\pm}}{\rm sen}(\delta_{l\pm})\rightarrow a_{l\pm} \ \ . 
\eeq

\section{loop integrals}

The basic loop integrals needed in order to perform the calculations of Fig. 2
are

\bea
&& I_{cc}^{\m\cdots} =  \int  \frac{d^4Q}{(2\pi)^4}\; 
\frac{\lp\frac{Q^\m}{\m}\cdots\rp}{[ (Q\!-\!q/2) ^2\! -\!\m^2] 
[(Q\!+\!q/2)^2 \!-\!\m^2]}
 \;,
\nonumber \\
&& \label{a`} \\
&& I_{\sb c}^{\m\cdots}= \int \frac{d^4Q}{(2\pi)^4}\;
\frac{\lp\frac{Q^\m}{\m}\cdots\rp}{[ (Q\!-\!q/2) ^2\! -\!\m^2] 
[(Q\!+\!q/2)^2 \!-\!\m^2]} \;
\frac{2m\m}{ [s -\mb^2 ] } \;. \nonumber \\
&&
\label{a2}
\eea

The integrals are dimensionless and have the following tensor structure
\bea
&& I_{cc} =  \frac{i}{(4\pi)^2} \lc \P_{cc}^{(000)}\rc \;,
\label{a3}\\
&& I_{cc}^{\m\n} = \frac{i}{(4\pi)^2}\lc \frac{q^\m q^\n}{\m^2} \; \P_{cc}^{(200)} 
+ g^{\m\n}\;\Pb_{cc}^{(000)}\rc \;,
\label{a4}\\
&& I_{\sb c} =  \frac{i}{(4\pi)^2}\lc \P_{\sb c}^{(000)} \rc\;,
\label{a5}\\
&& I_{\sb c}^{\m} =  \frac{i}{(4\pi)^2} \lc \frac{P^\m}{m} \; \P_{\sb c}^{(001)}\rc \;.
\label{a6}
\eea

Thus, the $\Pi$ integrals that appear in the text are
\bea
&& \P_{cc}^{(n00)} = - \int_0^1d a\;  (1/2-a)^n\; \ln \lp \frac{D_{cc}}{\m^2}\rp  \;,
\label{a7}\\[2mm]
&& \Pb_{cc}^{(000)}= - \;\frac{1}{2} \int_0^1 d a \; \frac{D_{cc}}{\m^2} \;
\ln \lp \frac{D_{cc}}{\m^2}\rp  \;,
\label{a8}\\[2mm]
&& \P_{\sb c}^{(00n)}=  \lp\!- 2m /\m \rp^{n+1} \int_0^1 d a\;a \int_0^1 d b \; 
\frac{\m^2\; (ab/2)^n}{D_{\sb c}}\;,  \nn \\
&&
\label{a9}
\eea

\ni
with
\bea
 D_{cc} &=& -a(1-a)\;q^2 + \m^2 \;,
\nn\\
 D_{\sb c} &=& -a(1-a)(1-b)\;q^2 \nn \\
&& + [\m^2 -ab\;(\m^2+m^2-\mb^2) + a^2b^2 \; m^2] \;.  
\nn
\eea



\begin{thebibliography}{99}

\bibitem{sw} L. Wolfenstein, Phys. Lett. {\bf 13}, 562 (1984).

\bibitem{KM} M. Kobayashi and T. Maskawa, Prog. Theor. Phys. {\bf 49}, 652
(1973).

\bibitem{pen} F. Gilman and M. Wise, Phys. Lett {\bf 93B}, 129 (1980).

\bibitem{WH} S. Weinberg, Phys. Rev. Lett. {\bf 37}, 657 (1976). 

\bibitem{dec} J. H. Christenson, J. W. Cronin, V. L. Fitch and R. Turlay,
Phys. Rev. Lett. {\bf 13}, 138 (1964). 

\bibitem{nir}  Y. Nir, hep-ph/0109090 (2001).

\bibitem{OK} S. Okubo, Phys. Rev. {\bf 109}, 984 (1958).

\bibitem{PA} A. Pais, Phys. Rev. Lett. {\bf 3}, 242 (1959).

\bibitem{don} J. F. Donogue, E. Golowich, W. A. Ponce and B. R. Holstein,
Phys. Rev. {\bf D21}, 186 (1980); J. F. Donogue, X. -G. He and S.
Pakvasa, Phys. Rev. {\bf D34}, 833 (1986); 
X. -G. He, H. Steger and G. Valencia, Phys. Lett. B {\bf 272}, 411 (1991);
O. E. Overseth and S. Pakvasa, Phys. Rev. {\bf 184}, 1163 (1969).


\bibitem{Kam} A. N. Kammal,  Phys. Rev. D {\bf 58}, 077501 (1998).

\bibitem{Dat} A. Datta, P. O'Donnell and S. Pakvasa ,  hep-ph/9806374 
(1998).

\bibitem{ta1} J. Tandean, A. W. Thomas and G. Valencia, Phys. Rev. D {\bf 64},
 014005 (2001).

\bibitem{meis} U. G. Mei$\beta$ner and J. A. Oller,  Phys. Rev. D {\bf 64}, 
014006 (2001).

\bibitem{ta2} J. Tandean, and G. Valencia, Phys. Rev. D {\bf 67}, 056001 
(2003).

\bibitem{xh1} X. G. He and G. Valencia,  Phys. Rev. D {\bf 52}, 5257 (1995).

\bibitem{xh2} X. G. He, H. Murayama, S. Prakvasa and G. Valencia ,  Phys. Rev. 
D {\bf 61}, 071701 (2000).

\bibitem{luk} K. B. Luk {\it et al.}, hep-ex/0005004.

\bibitem{ta3} J. Tandean, and G. Valencia, Phys. Lett. B  {\bf 451}, 382 
(1999).

\bibitem{PS} R. Kosh and E. Pietarinen, Nucl. Phys. {\bf A336}, 331 (1980).
 
\bibitem{LG} T. Becher and H. Leutwylwer, hep-ph/0103263 (2001).

\bibitem{PiN} E.T. Osypowski, Nucl. Phys. B {\bf 21}, 615 (1970);
M. G. Olsson and  E.T. Osypowski, Nucl. Phys. B {\bf 101}, 136 (1975);
H.T. Coelho, T.K. Das and M.R. Robilotta, Phys. Rev. C {\bf 28}, 1812 (1983).

\bibitem{LOI} B. Loiseau and S. Wycech, Phys. Rev. C  {\bf 63}, 034003 (2001).

\bibitem{keis} N. Keiser,  Phys. Rev. C {\bf 64}, 045204 (2001). 

\bibitem{BH} C. C. Barros and Y. Hama, Phys. Rev. C  {\bf 63}, 065203 (2001).

\bibitem{MAN} M. R. Robilotta, Phys. Rev. C  {\bf 63}, 044004 (2001).

\bibitem{CM} C. C. Barros and M. R. Robilotta, hep-ph/0209213, submitted to
physical review C.

\bibitem{LW} T. Becher and H. Leutwyler, Eur. Phys. Journal C {\bf 9}, 643 
(1999); JHEP {\bf 106}, 17 (2001).

\bibitem{EB}  H. B. Tang and P. J. Ellis, Phys. Rev. C {\bf 56}, 3363 
(1997).

\bibitem{PDG} Particle Data Group, C. Caso {\it et al}., Eur. Phys. J. C. 
{\bf 3}, 1 (1998).




\end{thebibliography}
\end{document}